\documentstyle[psfig]{l-aa}
\newcommand{\exo}{\hbox{EXOSAT}}
\newcommand{\mxu}{\hbox{4U\,1636$-$53}}
\newcommand{\ccd}{colour-colour diagram}
\newcommand{\hid}{hardness-intensity diagram}
\newcommand{\psm}{power spectrum}
\newcommand{\psa}{power spectra}
\newcommand{\xray}{\hbox{X-ray}\ }

\newcommand{\mdot}{$\dot M$}

\newcommand{\bhc}{black-hole candidate}
\newcommand{\lmxb}{low-mass \hbox{X-ray} binary}
\newcommand{\lmxbs}{low-mass \hbox{X-ray} binaries}
\begin{document}	

\title{Correlated \xray spectral and fast-timing behaviour of 4U\,1636$-$53}
\author{S.~Prins\inst{1} and M.~van der Klis\inst{1}}
\offprints{S.~Prins (saskia@astro.uva.nl)}
\institute{Astronomical Institute ``Anton Pannekoek'', University of
 Amsterdam,  
and
  Center for High Energy Astrophysics, \\
  Kruislaan 403, 1098 SJ Amsterdam, The Netherlands}
\thesaurus{06(02.01.2; 08.02.1; 08.09.2; 08.14.1; 13.25.5; 13.25.1)}
\date{Received , Accepted}
\maketitle
\markboth{S.~Prins \& M.~van der Klis: Correlated X-ray spectral and fast-timing behaviour of \mxu}{}

\begin{abstract}
We present a study of the persistent \xray emission from the \lmxb\
\mxu\ based on the entire archival EXOSAT ME data set.  We performed a
homogeneous analysis of all five EXOSAT observations in terms of the
correlated rapid X-ray variability and X-ray spectral properties by
means of power spectra, and colour-colour and hardness-intensity
diagrams, respectively.  Over the years we find similar patterns in
the \ccd\ and the \hid , but small shifts in their positions occur.
We find one case of an island-banana transition.  On all but one
occasion the differences in colours are smaller than 5~\%.
Between the two ``island state'' data sets we find a shift of 8 \% in
the soft colour, which must be at least partly intrinsic.

Clear correlations are present between the X-ray colours (which are
governed by the mass accretion rate) and the \psm .  With increasing
mass accretion rate the fractional rms amplitude of the power-law
shaped \psa l component that dominates the \psa\ below $\sim$\,1 Hz
rises, while the band-limited noise component at higher frequencies
drops.  The cut-off frequency of the band-limited component increases
when its amplitude decreases.  This is reminiscent of the behaviour of
\bhc s in the low state.


\end{abstract}
\section{Introduction}

The \lmxb\ \mxu\ was classified by Hasinger \& Van der Klis (1989,
hereafter HK89) as an atoll source.  The HK89 classification of
\lmxbs\ is based on the correlated variations of the \xray spectral
and rapid \xray variability properties.  HK89 distinguished two
(sub-)types of bright low mass \xray binaries, the Z sources and the
atoll sources, whose names were inspired by the shapes of the tracks
that they trace out in an \xray \ccd\ on time scales of hours to days.

Atoll sources are different from Z-sources in several ways: they are
less luminous, and the ``horizontal branch quasi-periodic
oscillations'' or ``normal/flaring branch quasi-periodic
oscillations'' seen in Z-sources have not been observed in them.  The
pattern traced out by an atoll source in a \ccd\ is usually (very)
roughly U-shaped.  The mass accretion rate (\mdot ) and (usually)
\xray count rate increase from left to right along the U.  At high
\mdot\ levels motion in the \ccd\ is relatively fast, and within
several hours an upwardly curved ``banana'' branch is traced out.  At
low \mdot\ levels, motion is slow, and observations lasting a few
hours to a few days produce ``islands'' in the \ccd --- this effect is
probably observational, and caused by a combination of relatively slow
motion in the \ccd\ and the limited length of the observations (HK89).
When an atoll source becomes very faint more complex patterns can
occur (\cite{1608},  \cite{michielb:1705}).

The \psa\ of atoll sources show two clear components (HK89), very low
frequency noise (VLFN), most prominent in the ``banana'' state, and
high frequency noise (HFN), most prominent in the ``island'' state.
VLFN has a power-law shape, HFN is a band-limited noise component with
a cut-off in the 5 -- 50\,Hz range.  A range of power spectral shapes
occurs, from a banana state power spectrum that consists of purely a
VLFN power law to an island state power spectrum that contains only a
strong HFN component with various mixed power-spectral shapes
containing both VLFN and HFN in between.  In recent work
(\cite{mut}) it has been suggested that the prominent
band-limited noise components in the power spectra of atoll sources in
the island state (atoll source HFN), \bhc s in the low state
(low-state noise) and Z-sources in the horizontal branch (Z-source
LFN) are all due to the same physical phenomenon, the accretion of
inhomogeneities (clumps) in the inner disk.  We will address this issue
in Section \ref{bhconnectie}.

\begin{table*}[t]
\caption[List of pointed observations of \mxu\ with the EXOSAT ME instrument]
{List of pointed observations of \mxu\ with the EXOSAT ME instrument.
In Columns 1 through 4 the observation start and end time and date are
given.
Column 5 gives the On Board Computer programs that were running,
Columns 7 and 9 give the time resolutions of the energy (7) and timing
(9) data respectively.  In Columns 6 and 8 a specification of the data
is given, Ar denotes that data from the argon chambers (1-20 keV)
only were used, Ar+Xe denotes that summed data from argon and xenon chambers
(5-50 keV) were used.  SB denotes one half on source, and one half on
the background, SS denotes that both halves were on source most of the
time (SSSS denotes that data were available for individual array
quadrants, and that all quadrants were on source most of the time).  In
Columns 10, 11, and 12 information about the available energy channels
for the HER data is listed: 10 begin channel, 11, end channel, 12
compression factor.  References in the last column refer to  
1)\,Damen et al., 1989, 1990, 2)\,\cite{HK89}, 3)\,\cite{lewin:bur},
4)\,\cite{vacca}, 5)\,\cite{leicester1}
}

\begin{tabular}{ll@{~~}r@{~~}rlclclcccl}
\hline
\multicolumn{13}{c}{EXOSAT ME observations of \mxu }\\
\hline
\multicolumn{4}{c}{Observation date}&
OBC & HTR & Time &HER & Time&\multicolumn{3}{c}{Start~~~End~~~Comp.}& Refs.\\ 
\multicolumn{4}{c}{and time (UT)}&programs&data&resolution & data &
resolution&  \multicolumn{3}{c}{chan.~~chan.~~factor}& \\
(1) & (2) &(3) & (4) &(5) & (6) &(7) & (8) &(9) & (10) &(11) & (12) &(13) \\
\hline
1983 &July &17~~17:57&18~~05:21&E4, E5, T3 &Ar   &7.8125 ms&SB  &0.3125 s&0 & 63& 2 &1,5	\\
1984 &May  & 6~~01:24& 6~~20:14&E5         & n/a & n/a     &SB  &0.3125 s&4 & 67& 2 &1, 4 \\
1984 &May  & 8~~11:04& 9~~10:00&E5, T3     &Ar   &7.8125 ms&SB  &0.3125 s&4 & 67& 2 &1, 4  \\
1984 &Sep  & 7~~05:15& 7~~14:41&E5, T3     & Ar  &7.8125 ms&SB  &0.625  s&4 & 67& 1 &1\\
1985 &Aug  & 6~~19:23& 7~~01:42&E5, T3     & Ar  &1.9531 ms&SS  &1.0    s&4 & 67& 1 &1, 2, 3\\
1985 &Aug  & 7~~11:10&10~~17:50&E5, T3     &Ar+Xe&7.8125 ms&SS  &0.3125 s&4 & 67& 2 &1, 2, 3\\
1985 &Aug  &11~~09:21&11~~10:50&E4, T3     &Ar+Xe&31.25  ms&SSSS&10     s&4 & 67& 1 & 1 \\
1985 &Sep  & 5~~06:20& 6~~05:56&E5, T3     &Ar+Xe&31.25  ms&SS  &0.3125 s&4 & 67& 1 &1\\
1985 &Sep  & 6~~16:16& 6~~20:10&E5, T3     &Ar+Xe& 31.25 ms&SS  &0.3125 s&4 & 67& 1 &1\\ 
\end {tabular}
\label{specificatie}
\end{table*}

Previous work on the EXOSAT data of \mxu\ included several studies
aimed at the burst properties (Turner \& Breedon, 1984, Lewin et al.,
1987, Damen et al., 1989, Damen et al., 1990),
and the \xray spectrum of the persistent emission (Breedon et al.,
1986, Vacca et al., 1987).
HK89 presented
one \ccd\ and three \psa\ based on part of the EXOSAT data. Van der
Klis et al.\ (1990) discussed the relation between the \ccd\ and the
burst properties.  In this paper we present a complete and homogeneous
analysis of the persistent emission in all EXOSAT data on \mxu , and
investigate in detail the correlation between the power spectra and
the \xray colours.

In Section \ref{observations} we describe the observations, in
Sections 3.1 -- 3.3 we describe the analysis of the spectral data in
terms of \ccd s and \hid s, and in Section \ref{pwspec} we describe
the analysis of the timing data in terms of \psa .  We discuss our
findings in Section \ref{discussie}.

\section{Observations}

\label{observations}
During the lifetime of EXOSAT \mxu\ was observed for a total of 176
hours with the Medium Energy (ME) instrument (\cite{exome}).  An
overview of the available ME data, and the time and energy resolution
for the different data sets can be found in Table \ref{specificatie}.
During the 1983 and 1984 observations, one half of the ME instrument
was pointed at the source, while the other half was slightly tilted to
monitor the background.  Typically every 4 hours swaps between the
halves were made.  During the 1985 observations, both halves were
pointed towards the source most of the time.  We removed all X-ray
bursts from the data using a table of burst onset times given by
Damen et al.~(1990).  For each burst, we
removed 260\,s of data, from 10\,s before to 250\,s after the onset
time.  For background determination, we used slew data or data from
offset array halves, whichever were closest in time.  We used,
whenever possible, only data from the argon detectors (1--20\,keV),
which have a better source to background count rate ratio than the
xenon detectors (5--50\,keV).  The spectral (HER) data were always from
the argon detectors, but in the case of the high-time resolution (HTR)
data, sometimes only summed argon plus xenon data were available.

\section{Analysis and results}
\subsection{Colour-colour and hardness-intensity diagrams}
\label{ccdhid}
We constructed \ccd s and \hid s using the argon-only HER data.
  Inspection of the spectrum of \mxu\ shows, that for energies above
$\sim$\,12 keV the spectrum is dominated by the background.  For the
\ccd s and \hid s we therefore only used the data below 11.55 keV.  We
divided these data into always exactly the same four broad energy
bands (see Table \ref{energies}).  To minimize the statistical
uncertainties, the energy bands were chosen such that each band had
roughly the same count rate.

\begin {figure*} 
\caption{Corrected X-ray \ccd s of \mxu .  Soft colour is defined as the ratio of the 2.4--3.4 keV
band to the 0.94--2.4 keV band, hard colour as the ratio of the
5.0--11.6 keV to the 3.4--5.0 keV band.  Each point represents a 200
second integration.  a) corresponds to July 1983, b) May 1984, c)
September 1984, d) August 1985, e) September 1985.  Typical error bars
are shown
}
\label{kleurkleur}
\end{figure*}

\begin {figure*}  
\caption{Corrected X-ray \hid s of \mxu\  for the different
data sets.  Hardness is defined as the ratio of the summed count rates
in the 3.4--11.5 keV band to the count rates in the 0.94--3.4 keV
band, intensity as the count rate (counts/sec/cm$^2$) in the 0.94--11.6 keV range.  Each point represents a 200 second integration.
a) corresponds to July 1983, b) May 1984, c) September 1984, d) August
1985, e) September 1985. Typical error bars
are shown  }
\label{hardness}
\end{figure*}

\begin{table}
\caption[Energy bands used for calculation of colours]
{The energy bands used for the calculation of colours.}
\begin{tabular}{ccr} 
\hline
Band&Start &End\\
&(keV)&(keV)\\
\hline
1&0.94 & 2.37 \\
2&2.37& 3.38\\
3&3.38& 4.96\\
4&4.96& 11.55\\
\end{tabular}
\label{energies}
\end{table}

\subsection{Instrumental corrections}
The detector gains varied slightly over the years, resulting in
changing channel energy boundaries, and the detectors were not
identical in gain and response.  Also, some of them broke down during
the life of \exo .  All this resulted in different effective
energy-channel boundaries in the on-board summed data from different
detectors.  We corrected for this by interpolating between the count
rates in the original energy channels according to the instantaneous
effective channel energy boundaries in such a way as to keep the
effective energy bands the same.  The count rates in each band were
corrected for deadtime (\cite{MEdeadtime}), and background; we always
used background count rates obtained from the same detectors as used
for the source observation.  We used these dead-time and background
corrected count rates to calculate the \xray colours.  The ``soft
colour'' is defined as the ratio of the count rate in band 2 to that
in band 1, the ``hard colour'' is defined as the ratio of band 4 to
band 3.  The ``hardness ratio'' used in the \hid s is the ratio of the
summed count rate in bands 3 and 4 to the summed count rate in bands 1
and 2, and the ``intensity'' is the sum of the count rates from bands
1, 2, 3, and 4.  We
uniformly used an integration time of 200 s per point.

To correct the colours and intensity for variation in the instrumental
response (as opposed to variations in effective energy boundaries), we
calculated \ccd s and \hid s of the Crab Nebula using the same energy
bands and analysis methods as for \mxu .  This procedure is described
in more detail in Kuulkers et al.~(1994).  We found that the
difference between the two halves of the \exo\ ME instrument caused by
the different detector responses is $\sim$1\,\% in the case of the
soft colours, and only 0.2\,\% for the hard colours.  From year to
year the changes in the colours of the Crab Nebula were usually about
0.5\,\%.  On two occasions we found larger deviations: $+$1.2\,\% in
soft colour and $-$1.3\,\% in hard colour in 1983, and $-$1.3\,\% in
soft colour in 1985.  The change in 1985 can be fully ascribed due to
the failure of detector 3.  The changes in intensity of the Crab
Nebula range from $+$4\,\% to $-$3\,\%.  We corrected the colours and
intensities of \mxu\ for these yearly variations and for the
differences in detector responses between the two halves by assuming
that the same effects as seen in the Crab Nebula also apply to \mxu .
Residual systematic effects can still be present, as the observations
of the Crab Nebula do not coincide in time with those of \mxu , and as
the spectral shapes are not the same, but based on previous experience
with this method (\cite{erik:gx5min1}) we expect these to be small.
The corrected \ccd s and \hid s of \mxu\ are shown in
Figs.~\ref{kleurkleur} and \ref{hardness}.

\begin{figure} 
\caption{Colour-colour diagram of the August 1985 data set showing the
regions adopted to define the rank numbers.}
\label{stukjes}
\end{figure}

\subsection{Motion in the colour-colour plane}

For each data set a pattern appears in the \ccd\ and \hid\ that
resembles those of other data sets.  However, even after the corrections
we made for instrumental effects, there is in general no exact
correspondence between the \ccd s and \hid s of the different
observations: there are shifts in the patterns from year to year.
Similar results were previously obtained for Z-sources
(\cite{erik:gx5min1}),
and in the following we assume that it is useful to describe the
behaviour of the atoll source \mxu\ in the colour-colour plane in
terms of a pattern with an approximately stable shape that is traced
out on time scales of hours to days, which moves through the plane on
longer time scales.   We will discuss below to what extent this motion
is intrinsic.  

\begin{figure}[t] 
\caption{Burst duration vs.~temperature for all bursts seen in
4U\,1636$-$53 with EXOSAT.  This figure is adapted from Fig.\,2 in
Damen et al.~(1989).  The filled symbols are for the September 1984
data }
\label{burstdurtemp}
\end{figure}

We used the \ccd\ for the August 1985 data set, which has the best
statistics and the greatest variety in source behaviour, to define the
motion in the colour-colour plane.  As can be seen in
Fig.~\ref{kleurkleur}, the only separate region in the \ccd\ of this
observation is the patch in the upper left section of the diagram,
which on the basis of its power spectrum, we identify as an island
(see below).   The gap between the island and the rest of the \ccd\ is 
observational; due to an Earth occultation no data were obtained for
eight hours, during which time the source moved considerably in the \ccd .

The only data set that matches the August 1985 \ccd\ without shifts is
the September 1985 one.  Shifts of $-$3\,\% in soft and $+$5\,\% in hard
colour are needed to fit the May 1984 observation, which clearly shows
a middle and upper banana, onto the August 1985 banana.  The intensity
in the 0.9 -- 11.6\,keV range of \mxu\ in this observation is $\sim$22\,\% lower than during the
corresponding parts of the August 1985 observation.  The July 1983
observation, which we identify as being entirely island-state, needs
an $+$8\,\% shift in soft colour and $+$2\,\% in hard colour to make it
correspond to the August 1985 island.  As previously reported
(\cite{leicester2}) the flux during this observation is high
and rising; it is 68\,\% -- 125\,\% higher than the
August 1985 island state intensity (in the 0.9 -- 11.6\,keV range).

For the brief September 1984 observation the state cannot be
ascertained purely on the basis of the morphology of the \ccd .
Therefore, we compared the properties of the three bursts seen in this
observation to those of the other bursts of \mxu\ using the results of
\cite{burstduration}, as burst duration and temperature are known to
correlate to the state \mxu\ (\cite{HKbursts})  is in.  To this purpose
we have reproduced Fig.~2 from Damen et al.~(1989) with different
symbols for different states and different observations (see
Fig.~\ref{burstdurtemp}).  From this comparison it is evident that the
September 1984 data correspond to the lower banana.  This implies a
shift of $-$4\,\% in soft colour and $+$4\,\%in hard colour, and
$-$19\,\% in intensity in the 0.9 -- 11.6\,keV range with respect to
the August 1985 data.

The observed changes in the colours and intensities of \mxu\ are much
larger than the observed changes in the colours and intensities of the
Crab Nebula.  Also, the direction of the changes in colours and
intensities of \mxu\ and the Crab Nebula  are not always the same.

\begin{figure} 
\caption{Colour-colour diagram of the May 1984 data with the rank
number regions superimposed.  Closed symbols refer to observations
where simultaneous HTR data are available, open symbols where HTR data
are missing. }
\label{maystukjes}
\end{figure}

For the purpose of quantifying the position of the source in the
pattern in the \ccd , we divided the colour-colour plane into eight
regions, numbered from 1 to 8 (see Fig.~\ref{stukjes}).  The August
1985 island corresponds to region 1, and the remaining wedge-like
regions divide the August 1985 banana into 7 sections, numbered 2 to
8.  The position of the source in the pattern is now defined by the
number of the region in the \ccd\ it is in after all the shifts needed
to make the pattern coincide with the August 1985 one were made.  We
shall refer to this number as the ``rank number'' from now on.  In this
way the May 1984 observation ranks 2 -- $\ga$ 8, the September 1984
observation 5, and the September 1985 observation 7 -- 8

\begin{figure*} 
\caption{The power spectra with best fits for the different rank
numbers in the \ccd s of August 1985.  The contribution of the
EXOSAT ME intrinsic band limited noise (Berger and Van der Klis, 1994)
is also plotted}
\label{augfits}
\end{figure*}

Only in the case of the July 1983 observation we think that its \ccd\
cannot be shifted onto the August 1985 \ccd, since we suggest it
corresponds to a state not present in the August 1985 observation (see
Section \ref{extremeisland}) .  We have divided the July 1983 data set
in two parts, corresponding to the two features in the \hid\ separated
at intensity 0.195 c/s/cm$^2$ (one part bent
upward to the left, the other part horizontal).  We have assigned
these rank numbers 0 and 0.5 .  In Fig.~\ref{burstdurtemp} it can be
seen that the bursts during this observation have the most extreme
values of all bursts of \mxu\ that were observed by EXOSAT.

\subsection{Power spectra}
\label{pwspec}

We calculated \psa\ from the 7.8\,ms and 1.9\,ms HTR data (see Table
\ref{specificatie}) using a time resolution of 7.8 ms throughout.  We
used data stretches with a length of 256\,s each containing 32768
points, so the resulting power spectra range from 1/256\,Hz to 64\,Hz.  All
incomplete time series (due to satellite telemetry drop outs, array
swaps or burst removal) were removed from the analysis.  This implied
a loss of 10--25\,\% of the data, depending on observation.  As longer
data stretches would have resulted in more data loss, we stuck with a
length of 256\,s and thus a lowest frequency of 0.0039\,Hz, even
though an extension to lower frequencies might have been of interest.
The 31.25\,ms data were analyzed separately: only in this case we 
calculated power spectra from data stretches of 1024\,s each containing 16384 points.

\begin{figure*} 
\caption{The power spectra for the different rank numbers in the \ccd s of July
1983, May 1984, September 1984, and September 1985.  The contribution
of the EXOSAT ME intrinsic band limited noise (Berger and Van der
Klis,1994) is also plotted}
\label{restfits}
\end{figure*}

\newcommand{\fix}{\multicolumn{1}{c}{--}}
\newcommand{\fixl}{\multicolumn{1}{c|}{--}}
\begin{table*}
\caption{List of fit parameters for power spectra of  \mxu . 
In Column 2 the rank number is given.  Columns 3 through 6 give 
VLFN fit results: 3 rms, 4 rms error, 5 power law index, 6 index error.  
Columns  7 through 12 give HFN fit results: 7 rms, 8 rms error, 
9 power law index, 10 index error, 11 cut-off frequency, 
12 frequency error.  
In Columns 13 and 14 we list the $\chi ^2$, and the degrees of freedom.
In Columns 15 and 16 we list the summed total count 
rate per sec and  summed back ground count rate per sec, as well as the ratio of number of detectors on 
source to total number of available detectors.  For all power spetral
fits the VLFN rms was integrated over 0.01 -- 1\,Hz.   For the
July 1983, May and September 1984, and August 1985 power spectral fits the HFN rms over 
 0.1 -- 64\,Hz.  Only for the September 1985 power spectral fits
the HFN rms was integrated over  0.1 -- 16\,Hz.   The listed errors are based
on a scan in $\chi ^2$ space using $\Delta\chi ^{2}=$1;  -- denotes that this parameter was kept fixed}
\begin{tabular}{ll|l@{~}rl@{~}r|r@{~}rl@{~}rr@{~}r|lcll}
\hline
\multicolumn{16}{c}{Fit parameters for power spectra of \mxu }\\
\hline
\multicolumn{1}{c}{Obs}&RN&\multicolumn{4}{|c|}{VLFN}&\multicolumn{6}{|c|}{HFN}& $\chi^2$&DOF&Count&Back\\
&& rms &error&index&error&rms&error&index&error&$\nu$$_{cut}$&error&&&rate&ground\\
&&\%&&&&\%&&&&(Hz)&&&&&\\
(1) & (2) &(3) & (4) &(5) & (6) &(7) & (8) &(9) & (10) &(11) & (12) &(13) &(14) &(15) &(16)\\
\hline
85    &  1 &  0  & \fix     & 1.30 & \fixl    &  7.84 & $+$\,0.75 &$-$0.30 &$+$\,0.19& 10.2 & $+$\,4.6&37.0 & 34 &   254.8  &   92.0  \\
Aug   &    &     &  \fix    &      & \fixl    &       & $-$0.76   &        &$-$0.27  &      & $-$3.0 &      &    & \multicolumn{2}{l}{~~8/8 Ar}\\   		
      &  2 &2.77 &$+$\,0.32 & 1.30 & \fixl    & 14.64 & $+$\,1.26 &$-$0.64 &$+$\,0.29& 28.9 &$+$25.3&49.7   & 33 &   794.0  &  620.0  \\
      &    &     &$-$0.37   &      & \fixl    &       & $-$1.37   &        &$-$0.40  &      &$-$10.1 &      &  & \multicolumn{2}{l}{~~8/8 Ar$+$Xe}\\
      &  3 &1.63 &$+$\,0.24 & 2.19 &$+$\,0.33 &  9.25 & $+$\,0.83 &$-$0.26 &\fix     & 100  &\fixl   & 29.9 & 34 &   848.7  &  620.0  \\
      &    &     &$-$0.23   &      &$-$0.28   &       & $-$0.92   &        &\fix     &      &\fixl   &      &  & \multicolumn{2}{l}{~~8/8 Ar$+$Xe}\\
      &  4 &0.94 &$+$\,0.17 & 1.30 & \fixl    &  4.45 & $+$\,1.09 &$-$0.26 &\fix     & 100  &\fixl   & 22.5 & 35 &   890.1  &  620.0  \\
      &    &     &$-$0.21   &      & \fixl    &       & $-$1.47   &        &\fix     &      &\fixl   &      &  & \multicolumn{2}{l}{~~8/8 Ar$+$Xe}\\
      &  5 &1.66 &$+$\,0.28 & 1.22 &$+$\,0.21 &  1.45 & $+$\,2.05 &$-$0.26 &\fix     & 100  &\fixl   & 28.7 & 34 &   928.5  &  620.0  \\
      &    &     &$-$0.29   &      &$-$0.17   &       & $-$4.29   &        &\fix     &      &\fixl   &      &  & \multicolumn{2}{l}{~~8/8 Ar$+$Xe}\\
      &  6 &2.31 &$+$\,0.17 & 1.33 &$+$\,0.11 &  2.04 & $+$\,1.49 &$-$0.26 &\fix     & 100  &\fixl   & 33.1 & 34 &   975.1  &  620.0  \\
      &    &     &$-$0.17   &      &$-$0.10   &       & $-$4.07   &        &\fix     &      &\fixl   &      &  & \multicolumn{2}{l}{~~8/8 Ar$+$Xe}\\
      &  7 &2.50 &$+$\,0.14 & 1.34 &$+$\,0.08 &  3.08 & $+$\,1.03 &$-$0.26 &\fix     & 100  &\fixl   & 31.0 & 34 &  1007.7  &  620.0  \\
      &    &     &$-$0.14   &      &$-$0.07   &       & $-$1.63   &        &\fix     &      &\fixl   &      &  & \multicolumn{2}{l}{~~8/8 Ar$+$Xe}\\
      &  8 &2.30 &$+$\,0.15 & 1.42 &$+$\,0.09 &$-$1.36& $+$\,3.80 &$-$0.26 &\fix     & 100  &\fixl   & 35.4 & 34 &  1032.8  &  620.0  \\
      &    &     &$-$0.15   &      &$-$0.09   &       & $-$1.74   &        &\fix     &      &\fixl   &      &  & \multicolumn{2}{l}{~~8/8 Ar$+$Xe}\\
83    &0.0 &  0  & \fix     & 1.30 & \fixl    & 10.53 & $+$\,0.94 &$-$0.21 &$+$\,0.19& 12.9 &$+$\,6.4& 23.4 & 34 &   224.6  &   92.0  \\
Jul   &    &     & \fix     &      & \fixl    &       & $-$0.97   &        &$-$0.26  &      &$-$4.0  &      &  & \multicolumn{2}{l}{~~4/8 Ar}\\
      &0.5 &1.45 &$+$\,0.18 & 1.30 & \fixl    &  9.63 & $+$\,0.71 &$-$0.49 &$+$\,0.29& 18.9 &$+$10.8&40.8 & 33 &   262.0  &   92.0  \\
      &    &     &$-$0.20   &      & \fixl    &       & $-$0.76   &        &$-$0.41  &      &$-$5.7  &      &  & 	\multicolumn{2}{l}{~~4/8 Ar}\\
84    &2.75&1.28 &$+$\,0.31 & 1.30 & \fixl    &  7.92 & $+$\,1.52 &$-$0.26 &\fix     & 100  &\fixl   & 33.2 & 35 &   190.9  &   92.0  \\
May   &    &     &$-$0.41   &      & \fixl    &       & $-$1.90   &        &\fix     &      &\fixl   &      &  & \multicolumn{2}{l}{~~4/8 Ar}\\
      &3.25&1.47 &$+$\,0.22 & 1.30 & \fixl    &  2.91 & $+$\,2.34 &$-$0.26 &\fix     & 100  &\fixl   & 33.4 & 35 &   193.7  &   92.0 \\
      &    &     &$-$0.25   &      & \fixl    &       & $-$6.17   &        &\fix     &      &\fixl   &      &  &    \multicolumn{2}{l}{~~4/8 Ar}\\
      &3.75&1.43 &$+$\,0.21 & 1.30 & \fixl    &$-$3.56& $+$\,6.05 &$-$0.26 &\fix     & 100  &\fixl   & 34.6 & 35 &   199.0  &   92.0 \\
      &    &     &$-$0.24   &      & \fixl    &       & $-$2.06   &        &\fix     &      &\fixl   &      &  & 	\multicolumn{2}{l}{~~4/8 Ar}\\
      &4.25&0.72 &$+$\,0.37 & 1.30 & \fixl    &  6.01 & $+$\,1.78 &$-$0.26 &\fix     & 100  &\fixl   & 45.5 & 35 &   204.5  &   92.0  \\
      &    &     &$-$1.12   &      & \fixl    &       & $-$2.62   &        &\fix     &      &\fixl   &      &  & 	\multicolumn{2}{l}{~~4/8 Ar}\\
84    &  5 &1.73 &$+$\,0.28 & 1.45 &$+$\,0.25 &  5.62 & $+$\,1.11 &$-$0.26 &\fix     & 100  &\fixl   & 42.3 & 34 &   213.0  &   92.0   \\
Sep   &    &     &$-$0.28   &      &$-$0.20   &       & $-$1.39   &        &\fix     &      &\fixl   &      &  & \multicolumn{2}{l}{~~4/8 Ar}\\
85    &  7 &2.57 &$+$\,0.16 & 1.33 &$+$\,0.11 &  1.86 & $+$\,0.74 &$+$\,0.17&$+$\,0.30&10000&\fixl   & 17.8 & 29 &   890.5  &  541.0  \\
Sep   &    &     &$-$0.22   &      &$-$0.08   &       & $-$1.63   &        &$-$0.44  &      &\fixl   &      &  & \multicolumn{2}{l}{~~7/7 Ar$+$Xe}\\
      &  8 &2.11 &$+$\,0.19 & 1.53 &$+$\,0.16 &  2.58 & $+$\,0.54 &$+$\,0.35&$+$\,0.18&10000&\fixl   & 33.0 & 29 &   906.0  &  541.0  \\
      &    &     &$-$0.23   &      &$-$0.12   &       & $-$0.70   &        &$-$0.21  &      &\fixl   &      &  &    \multicolumn{2}{l}{~~7/7 Ar$+$Xe}\\
\end{tabular}
\label{pwtabel}
\end{table*}

We subtracted the predicted Poisson level as
altered by dead time processes from each power spectrum using
expression 3.9 from Van der Klis (1989) with an
effective deadtime of 10.6 $\mu$s (\cite{michielb:cygx3}), and then
averaged the (Leahy-normalized) power spectra according to rank
number. Usually, more than 50 power spectra were averaged.  Power
spectra of time series which overlap with more than one region in the
\ccd\ contribute to the average proportionally to the fractional
extent of the overlap.  

We fitted the \psa\ using a function consisting of two components,
very low frequency noise (VLFN) described by a power law, and high
frequency noise (HFN) described by a power law with an exponential
cut-off (HK89).  In a number of cases we found that the five-parameter
fit function was not sufficiently constrained by the data to determine
all parameters.  In particular, the HFN cut-off frequency was usually
not well-constrained when it was near or beyond the Nyquist frequency.
Also, the HFN power-law index was in those cases usually hard to
measure.  For rank numbers $\ga$\,4 the HFN component was hardly
measurable above the EXOSAT ME intrinsic band limited noise reported
by Berger and Van der Klis (1994).  Therefore, for rank numbers $>$\,2
we fixed the HFN cut-off frequency and the power-law index at the
values found by Berger and Van der Klis (1994) for the intrinsic
noise, leaving only the strength of the HFN as a fit parameter.
For most power spectra for rank numbers $\la$ 4 the VLFN power-law
index was difficult to measure.  In those cases we fixed the power-law
index at a value representative of that seen in other power spectra.

For the September 1985 power spectra the Nyquist frequency is only
16\,Hz. Therefore, we only tried to measure the strength of the HFN by
fixing the cut-off frequency at a  high value.

To all fractional rms amplitudes derived from the fits we applied a
correction factor (B + S)/S, where B is the background count rate and
S the source count rate, so as to obtain the fractional rms amplitude
of the source flux.  This background correction factor amounts to
$\sim$\,1.54 to $\sim$\,1.93 for the data that consisted of argon only,
and $\sim$\,2.48 to $\sim$\,4.56 for the data that consisted of argon
plus xenon (due to the higher background for the xenon detectors).  We
corrected the rms values for the presence of the EXOSAT ME
intrinsic band limited noise reported by Berger and Van der Klis
(1994).

For presentation purposes, we normalized the average power spectra
according to the (rms/mean)$^2$/Hz normalization (see Van der Klis
(1995) for the formulae used), so that the
integrated power directly corresponds to the source fractional rms
amplitude squared.  The power spectra corresponding to the various
colour-colour regions of the August 1985 observation are shown in
Fig.~\ref{augfits}; the power spectra of the other observations in
Fig.~\ref{restfits}.  The best fit to the power spectrum, and the
contribution of the EXOSAT ME instrumental component, as well as date,
rank number and detectors used (argon or argon plus xenon) are shown
in each diagram.

In Table \ref{pwtabel} we list all the fit results with errors based
on a scan in $\chi ^2$ space using $\Delta\chi ^{2}=$1.  The reason
that the May 1984 observations in Table \ref{pwtabel} only cover rank
numbers 2 -- 5, whereas its \ccd\ clearly extends to the upper banana
(even beyond rank number 8), is that for the HER data obtained for
ranks $\la$ 6 there were no simultaneous HTR data (see
Fig.~\ref{maystukjes}).  For all power spectral fits the VLFN rms was
integrated over 0.01 -- 1\,Hz.  For the July 1983, May and September
1984, and August 1985 power spectral fits the HFN rms was integrated
over 0.1 -- 64\,Hz.  Only for the September 1985 power spectral fits
the HFN rms was integrated over 0.1 -- 16\,Hz.

\begin{figure} 
\caption{VLFN and HFN fractional rms amplitudes vs.~rank number.
Closed symbols refer to argon plus xenon data, open symbols to
argon-only data.}
\label{vlfnhfn}
\end{figure}

\section{Discussion}
\label{discussie}
The analysis described in the previous section now enable us to answer
the question to what extent the position in the pattern in the
colour-colour diagram completely determines the character of the rapid
\xray variability.  In Fig.~\ref{vlfnhfn} the fractional rms
amplitudes of the VLFN and
HFN are plotted as a function of  rank number.  Open symbols refer to
argon-only data, whereas filled symbols are for argon plus xenon data.
There is a clear trend of the HFN rms dropping with rank, from as high
as 10--15\,\% at rank 1 and 2 to consistent with zero in the upper
banana.  The VLFN  is weak ($\sim$\,1\,\%) at low rank and increases
to $\sim$\,2.5\,\% in the upper banana.  Comparing the argon and argon
plus xenon data, there is no strong evidence for a simple photon
energy dependence, except perhaps for a slightly lower HFN in argon
plus xenon in the lower ranked regions.

The correspondence between the various observations is good:  the same
trends are seen in independent data sets.  When comparing the
argon-only and the argon plus xenon data separately, there are no
significant deviations from a smooth dependence of rms on rank in
either VLFN or HFN.  The HFN cut-off frequency (not plotted) was only
measurable for ranks < 3, where on two different occasions it
seemed to increase from $\sim$\,10 to $\sim$\,25\,Hz. 
There is no strong evidence for a dependence of the VLFN or HFN
power-law  indices with rank number.  

The 1983 island state power spectra both have a higher HFN rms than
the 1985 island state power spectrum.  As the 1983 island state is
different from the 1985 island state in terms of the burst behaviour
(see Fig.~\ref{burstdurtemp}), we believe the colour difference may be
intrinsic as well.  \label{extremeisland} In 4U~1608$-$52
(\cite{1608}) and 4U~1705$-$44 (\cite{michielb:1705}) the most extreme
island behaviour occurs at the upper right of the \ccd, with less
extreme island behaviour occurring at the lower left of the \ccd.  We
suggest that the 1983 island state is a case of an extension to the
island state as described by HK89 in the direction of the (hard) upper
right towards extreme island behaviour.
Hence, the difference in the \ccd\ of 1983 should be primarily
ascribed to a different state, and not to a  change of the
position of the island state pattern in the \ccd .  
The observed changes in
the other data sets are also larger than the first order correction we
applied, but systematic effects may still be present.  

In the case of the August 1985 observation we find a high value for
the HFN and VLFN rms at rank number 2, i.e., immediately after the
observational gap between the island and lower banana state.  In
August 1985 we measure a higher HFN rms at rank number 2, than for
(island) rank numbers 1 (1985) and 0 and 0.5 (July 1983).  A caveat is
that the rms values for rank numbers 0 -- 1 were all obtained from
argon-only data, whereas those for the August 1985 rank numbers 2 -- 8
were from combined argon and xenon data.  For rank numbers > 2 the HFN
rms rapidly decreases to a level consistent with zero.  At rank number
2 there is still a measurable HFN cut-off frequency, which is clearly
higher than for the island state.  For rank numbers 2 -- 4 the VLFN
rms decreases, after which it rises again for rank numbers 4 -- 7 ,
with perhaps a flattening for rank number 8.  The VLFN and HFN rms
values obtained for the September 1985 observation (rank number 7 --
8) are fully consistent with those of rank number 7 -- 8 for the
August 1985 observation.

Although the transition from low state to high state is not as gradual
as had previously been suggested by HK89, their general ideas (--the
spectral and the timing properties are correlated and governed by one
parameter, the instantaneous mass accretion rate, --the HFN decreases
in strength from low (island) state to high (banana) state, --the VLFN increases in
strength from low state to high state) still hold.  
It is not clear
however, how the non-monotonicity in the HFN and VLFN can be explained.
As yet it is unclear what causes the observed high rms in both the HFN
and VLFN at the
transition from the island to the banana state.  

\label{bhconnectie}
In a recent paper Van der Klis (1994) argues that the HFN of atoll
sources is similar to the noise of black hole candidates in the low
state, while very different from the HFN of Z-sources.  This indeed
seems to be the case, at least for the low (island) state: in \mxu\
the low state HFN decreases while its cut-off frequency increases,
just as is observed from the band-limited noise in black hole
candidates (Belloni \& Hasinger, 1990, Van der Klis, 1994, M\'endez \&
Van der Klis, 1996).  The only atoll source in which this has been
observed so far is 4U\,1608$-$52 (\cite{1608}), but in that case at a
much stronger fractional rms.

\acknowledgements
We thank Erik Kuulkers for assistance with the corrections based on
Crab data, and John
Telting for comments on an earlier version of the manuscript.
This was supported in part by the Netherlands Organisation for
Scientific Research (NWO) under grant number PGS 78-277.  

\end{document}